# Reevaluation of the Hadronic Contribution to $\alpha(M_Z^2)$ (revised) [⋆]

Morris L. Swartz

*Stanford Linear Accelerator Center*

*Stanford University, Stanford, California, 94309*

## ABSTRACT

We reevaluate the hadronic part of the electromagnetic vacuum expectation value using the standard dispersion integral approach that utilizes the hadronic cross section measured in $e^+e^-$ experiments as input. Previous analyses are based upon point-by-point trapezoidal integration which has the effect of weighting all inputs equally. We use a technique that weights the experimental inputs by their stated uncertainties, includes correlations, and incorporates some refinements. We find the hadronic contribution to the fractional change in the electromagnetic coupling constant at $q^2 = M_Z^2$, $\Delta\alpha(M_Z^2)$, to be $0.02666 \pm 0.00075$, which leads to a value of the electromagnetic coupling constant, $\alpha^{-1}(M_Z^2) = 129.08 \pm 0.10$. This value significantly shifts the Standard Model predictions for the effective weak mixing angle measured at the $Z$ pole and moderately shifts the predicted $Z$ width.

Submitted to *Physical Review D*

---

⋆ Work supported by the Department of Energy, contract DE–AC03–76SF00515.

## 1. Introduction

At the current time, a large program of precise electroweak measurements is being conducted throughout the world. The object of this program is to test the electroweak Standard Model by comparing the measured values of a large set of electroweak observables with the predictions of the Minimal Standard Model (MSM). The Standard Model calculations have been performed to full one-loop accuracy and partial two-loop precision by a large community of researchers. In all of these calculations, it is necessary to evaluate the one-particle-irreducible contributions to the photon self-energy $\Pi_{\gamma\gamma}(q^2)$ or the related quantity $\Pi'_{\gamma\gamma}(q^2) \equiv (\Pi_{\gamma\gamma}(q^2) - \Pi_{\gamma\gamma}(0))/q^2$ at the $Z$ mass scale $q^2 = M_Z^2$. These quantities are usually absorbed into the definition of the running electromagnetic coupling $\alpha(q^2)$,

$$\alpha(q^2) \equiv \frac{\alpha_0}{1 - \Pi'_{\gamma\gamma}(q^2)}, \tag{1}$$

where $\alpha_0 = 1/137.0359895(61)$ is the electromagnetic fine structure constant. This quantity is also represented as the fractional change in the electromagnetic coupling constant $\Delta\alpha$,

$$\Delta\alpha(q^2) = \frac{\alpha(q^2) - \alpha_0}{\alpha(q^2)} = \Pi'_{\gamma\gamma}(q^2). \tag{2}$$

Using analytic techniques and the optical theorem applied to the amplitude for s-channel Bhabha scattering, the quantity $\Delta\alpha$ has been related to the cross section for the process $e^+e^- \to \gamma^* \to$ all ($\sigma_{tot}$) as follows,[1]

$$\Delta\alpha(q^2) = \frac{\alpha_0}{3\pi} \mathrm{P} \int_{4m_\pi^2}^{\infty} ds \frac{q^2}{s(q^2 - s)} R_{tot}(s), \tag{3}$$

where $R_{tot}(s) \equiv \sigma_{tot}(s)/\sigma_{\mu\mu}(s)$ is the ratio of the total cross section to the (massless) muon pair cross section at the center-of-mass energy $\sqrt{s}$. It should be noted in passing that equation (3) is correct to all orders in $\alpha_0$ and relies only upon the assumption that the real part of $\Pi_{\gamma\gamma}$ is much larger than its imaginary part (the next-order correction is proportional to $\text{Im}^2\Pi_{\gamma\gamma}/|\Pi_{\gamma\gamma}|^2$ which is approximately $3 \times 10^{-4}$ at $q^2 = M_Z^2$). It is straightforward to evaluate equation (3) for the continuum leptonic cross sections.[2] In the limit that the scale $q^2$ is much larger than the square of the lepton mass $m_\ell^2$, the contribution of the continuum leptonic cross sections is given by the following expression,

$$\Delta\alpha_\ell(q^2) = \frac{\alpha_0}{3\pi}\sum_\ell \left[-\frac{5}{3} + \ln\frac{q^2}{m_\ell^2}\right].\tag{4}$$

The remaining contributions to $R_{tot}$ consist of the continuum hadronic cross section and the $J^P = 1^-$ resonances and are labelled $R_{had}$. Since the cross sections for these final states are not accurately calculable from first principles, experimental inputs are used to evaluate the remaining portion of equation (3),

$$\Delta\alpha_{had}(q^2) = \frac{\alpha_0}{3\pi}\text{P}\int\limits_{4m_\pi^2}^{\infty} ds\frac{q^2}{s(q^2-s)}R_{had}(s).\tag{5}$$

Equation (5) has been evaluated at the $Z$ boson mass scale several times.[3,4,5] A complete description was given by Burkhardt, Jegerlehner, Penso, and Verzegnassi[4] in 1989. The result was updated by Jegerlehner[5] in 1991 (to include measurements from the Crystal Ball Collaboration),

$$\Delta\alpha_{had}(M_Z^2) = \begin{cases} 0.0288 \pm 0.0009, & \text{Reference 4} \\ 0.0282 \pm 0.0009, & \text{Reference 5.} \end{cases}\tag{6}$$

The authors of Reference 4 perform the integration in three parts: the hadronic continuum above $\sqrt{s} = 1$ GeV, the $\pi^+\pi^-$ final states above threshold; and the $\omega$, $\phi$, $\psi$, and $\Upsilon$ resonances. The continuum integration is performed by linearly interpolating between the data points. The resonance contributions were calculated from an analytic expression which results from integrating a Breit-Wigner lineshape and depends upon the masses, widths, and leptonic widths of each resonance. The uncertainties on each contribution were estimated by techniques which appear to be very conservative. The work reported in this paper was undertaken initially to estimate a more accurate error. We have indeed performed what we believe to be a more accurate analysis, however, it is our conclusion that the uncertainty estimated by the authors of Reference 4 is not overestimated.

## 2. The Analysis

Any attempt to combine the results of many experiments is a perilous undertaking. Many different techniques and approaches have been used. Not all researchers have addressed all possible problems nor are systematic error estimates performed in uniform ways or to uniform standards. We therefore adopt some the techniques of the Particle Data Group.[6] Older measurements which are contradicted by newer, more precise work are excluded from the analysis. Parameter uncertainties that are extracted from fits with $\chi^2$ per degree of freedom (dof) larger than one are rescaled by the factor $\sqrt{\chi^2/dof}$.

### 2.1 THE DATA

The approach to the evaluation of equation (5) is driven by the form of the data themselves. The total hadronic cross section can be decomposed into four pieces: the hadronic continuum above $W \equiv \sqrt{s} = 1$ GeV, the charged two-body final states $\pi^+\pi^-$ and $K^+K^-$ from their respective thresholds to 2.6 GeV, and hadronic resonances (excluding charged two-body final states). Since equation (5) is linear in the hadronic cross section, we decompose $\Delta\alpha_{had}$ as follows,

$$\Delta\alpha_{had}(q^2) = \Delta\alpha_{had}^{\text{cont}}(q^2) + \Delta\alpha_{had}^{\pi^+\pi^-}(q^2) + \Delta\alpha_{had}^{K^+K^-}(q^2) + \Delta\alpha_{had}^{\text{res}}(q^2), \qquad (7)$$

where the four terms on the right-hand side correspond to the four pieces of the hadronic cross section.

The rationale for this decomposition is as follows. The region below $W = 1$ GeV is dominated by the $\rho$, $\omega$, and $\phi$ resonances. The electromagnetic form factors for the processes $e^+e^- \to \pi^+\pi^{-\,[7-14]}$ and $e^+e^- \to K^+K^{-\,[14-18]}$ are measured well from threshold to $W \simeq 2$ GeV. Resonances do not account for all of the $\pi^+\pi^-$ and $K^+K^-$ cross section in this region. On the other hand, essentially all other two-body and three-body final states are associated with the resonances. Measurements of three-pion final states near $W = 1$ GeV[19] show the non-resonant portion to be consistent with zero. Similarly, measurements of various two-body final states such as $K_L^0 K_S^0$ show small non-resonant cross sections.[16] The cross sections for four-pion final states become significant above 1 GeV but are small below that energy.[20] The $\gamma\gamma2$ experiment[21] at the ADONE storage ring at Frascati has measured the hadronic cross section ratio for three or more hadron final states, $R_{had}^{\geq 3}$ from $W = 1.42$ GeV to $W = 3.09$ GeV. They have also presented several points from 1 GeV to 1.4 GeV that are composed of various multipion cross sections from Novosibirsk and Orsay[20,19,22] and are claimed to approximate $R_{had}^{\geq 3}$. Measurements beginning at $W = 2.6$ GeV by the MARK I,[23] DASP,[24] and PLUTO[25] Collaborations claim to measure the entire cross section. We therefore conclude that $R_{had}$ is well approximated below $W_1 = 2.6$ GeV by a sum of the $\pi^+\pi^-$ and $K^+K^-$ contributions from threshold to $W_1$ (where they are much smaller than $R_{had}^{\geq 3}$); the $R_{had}^{\geq 3}$ measurements from 1 GeV to $W_1$; and the $\rho$, $\omega$ and $\phi$ resonances where the hadronic widths are adjusted to remove the $\pi^+\pi^-$ and $K^+K^-$ final states that are already included explicitly. Note that the several broad $e^+e^-$ resonances between the $\phi(1020)$ and $W = 2$ GeV are implicitly contained in the two-body or $R_{had}^{\geq 3}$ categories. Since the $\pi^+\pi^-$ and $K^+K^-$ cross sections are very small at $W_1$, the $R_{had}^{\geq 3}$ and total continuum $R_{had}$ measurements should be continuous at this point.

At center-of-mass energies larger than $W_1$, many measurements of the hadronic continuum and resonances exist. The region of the charm threshold from $W = 3.6$ GeV to $W = 5.0$ GeV is complicated and not well measured. The MARK I, DASP, and PLUTO Collaborations all observe an enhancement beyond the expected threshold shape. The DASP data show three resolved resonances. The MARK I and PLUTO data are consistent with the DASP data but do not cleanly resolve the resonances. We choose to follow the Particle Data Group and recognize the DASP resonances: $\psi(4040)$, $\psi(4160)$, and $\psi(4415)$. The $\psi$ family therefore consists of six states.

Between 5 GeV and 10.4 GeV, the MARK I, DASP, PLUTO, Crystal Ball,[26] LENA,[27] CLEO,[28] CUSB,[29] and DESY-Heidelberg[30] Collaborations have published $R_{had}$ measurements which are are plotted in Figure 1. The error bars include only point-to-point uncertainties. The recently published Crystal Ball measurements have a systematic normalization uncertainty of 5.2%. The other measurements have normalization uncertainties in the range 6.8-10%. The data are also compared with the recent QCD prediction of Chetyrkin and Kuhn[31] which includes quark mass effects. At $W = 5$ GeV, the MARK I data are consistent with other measurements. As $W$ increases, they show a systematic increase in $R_{had}$ and suggest the presence of a structure near 6.6 GeV. Including the quoted 10% normalization uncertainty, the MARK I data are larger than the more precise measurements by approximately two standard deviations. The reader is reminded that first generation detectors like MARK I, DASP, and PLUTO were small acceptance devices that necessarily involved large acceptance corrections without the benefit of good event structure modelling. After acceptance corrections and a $\tau$-lepton subtraction, the MARK I group observed that two-charged-prong events constituted nearly 20% of the hadronic cross section of $R$ at $W = 7$ GeV. This is about 1.5 times the two-prong rate due to $\tau^+\tau^-$ final states and three times the rate that is predicted[32] by the JETSET 7.3 Monte Carlo program.[33] While this may not be wrong, we choose to exclude data from the first generation experiments when more modern results are available. Such data are available above charm threshold. Unfortunately, we are constrained to use very old continuum measurements below charm threshold.

The Particle Data Group lists six $\Upsilon$ family resonances between 9.4 GeV and 11 GeV. All are included in the resonance contribution.

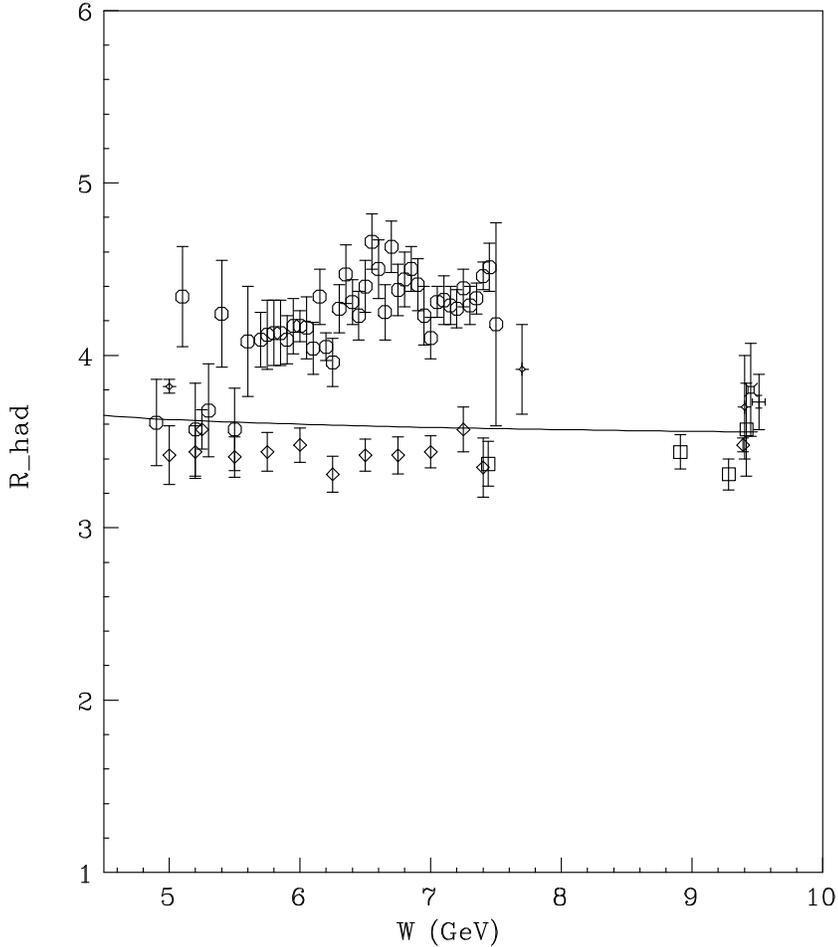

Figure 1. The $R_{had}$ measurements of the MARK I[23] (circles), PLUTO[25] (horizontal marks), Crystal Ball[26] (diamonds), LENA[27] (squares), CLEO[28] (stars), CUSB[29] (X's), and DESY-Heidelberg[30] (vertical crosses) Collaborations in the region between $W = 5$ GeV and $W = 9.4$ GeV. The error bars include point-to-point uncertainties only. A recent QCD calculation[31] which includes quark mass effects is shown as a solid line for $\alpha_s(M_Z^2) = 0.125$.

Above b-quark threshold, a number of $R_{had}$ measurements have been carried out by the PEP and PETRA experiments.[34-39] However at energies above $W = 34$ GeV, $Z$-$\gamma$ interference becomes significant. We therefore use only those measurements in the region $W \leq 34$ GeV where the correction for electroweak interference is less than 1%.

We expect that $R_{had}$ is well described by perturbative QCD in the region above b-quark threshold. This implies that the measurement of the strong coupling constant, $\alpha_s(M_Z^2)$, extracted from a fit to the lineshape parameters of the $Z$ boson by the LEP Collaborations[40]

$$\alpha_s(M_Z^2) = 0.125 \pm 0.005, \tag{8}$$

provides a precise measurement of $R_{had}$ at $W = M_Z$. We have verified that the value of $\alpha_s(M_Z^2)$ given in equation (8) is insensitive to the value of the electromagnetic coupling constant $\alpha(M_Z^2)$ used in the fitting procedure ($\alpha(M_Z^2)$ can be left as a free parameter with essentially no effect on the extracted value of $\alpha_s$). To convert $\alpha_s(M_Z^2)$ into a determination of $R_{had}(M_Z)$, we use the third-order $\overline{\text{MS}}$ QCD expression[41]

$$R_{QCD}(s) = 3 \sum_f Q_f^2 \beta_f \frac{(3 - \beta_f^2)}{2}$$
$$\cdot \left\{ 1 + \left[ \frac{\alpha_s(s)}{\pi} \right] + r_1 \left[ \frac{\alpha_s(s)}{\pi} \right]^2 + r_2 \left[ \frac{\alpha_s(s)}{\pi} \right]^3 \right\}, \tag{9}$$

where: $Q_f$ is the final state fermion charge, $\beta_f = \sqrt{1 - 4m_f^2/s}$ is the fermion velocity in the $e^+e^-$ center-of-mass frame ($m_f$ is the fermion mass), and the coefficients are functions of the number of active flavors $N_f$,

$$r_1 = 1.9857 - 0.1153 N_f$$
$$r_2 = -6.6368 - 1.2002 N_f - 0.0052 N_f^2 - 1.2395 \frac{\left( \sum Q_f \right)^2}{3 \sum Q_f^2}. \tag{10}$$

The resulting value of $R_{had}(M_Z)$ is,

$$R_{had}(M_Z) = 3.818 \pm 0.006. \tag{11}$$

The following three sections of this chapter describe the evaluation of: the continuum contribution $\Delta\alpha_{had}^{\text{cont}}$, the contributions of the charged two-body final states $\Delta\alpha_{had}^{\pi^+\pi^-}$ and $\Delta\alpha_{had}^{K^+K^-}$, and the resonance contribution $\Delta\alpha_{had}^{\text{res}}$.

## 2.2 THE HADRONIC CONTINUUM

The authors of Ref. 4 evaluated equation (5) for the continuum contribution by performing a trapezoidal integration with measured values of $R_{had}$. Their approach has two advantages: it is unbiased by human prejudice about the functional form of $R_{had}(s)$, and it would automatically account for undiscovered resonances which are broad as compared with the spacing of measurements. Unfortunately, this technique also has a serious shortcoming: it does not take experimental errors into account properly. All data points receive equal weight irrespective of their experimental precision. An experiment which publishes a large number of imprecise data points receives more weight than an experiment which publishes fewer precisely measured ones.

We avoid this problem by fitting the data to an appropriate functional form $R_{fit}(s; a_k)$ where $a_k$ are the parameters of the function. In the absence of undiscovered resonances, $R_{had}$ can be described by a continuous function. A $\chi^2$ fit has the virtue that measurements are weighted by their experimental errors. To do this, we make the conservative assumption that all normalization uncertainties within an appropriate grouping of measurements are 100% correlated. The $\chi^2$ function therefore has the form,

$$\chi^2 = \sum_{i,j} \Big[ R_{had}^i - R_{fit}(s_i; a_k) \Big] \mathcal{W}_{ij} \Big[ R_{had}^j - R_{fit}(s_j; a_k) \Big], \tag{12}$$

where $R_{had}^i$ is the value of $R_{had}$ measured at energy $s_i$ and the inverse elements of the weight matrix $\mathcal{W}_{ij}$ are given by the following expression,

$$[\mathcal{W}^{-1}]_{ij} = \begin{cases} \sigma_i^2(\text{ptp}) + \sigma_i^2(\text{norm}), & i = j \\ \sigma_i(\text{norm})\sigma_j(\text{norm}), & i \neq j, \text{ same grouping} \\ 0, & i \neq j, \text{ different grouping} \end{cases} \tag{13}$$

where $\sigma_i(\text{ptp})$ and $\sigma_i(\text{norm})$ are the point-to-point (statistical and systematic) and normalization uncertainties associated with the $i^{th}$ measurement.

Equation (5) is evaluated by performing a Simpson's Rule integration using the function $R_{fit}$ and the best estimate of the parameters. The parameter uncertainties $\delta a_k$ reflect the point-to-point and normalization uncertainties to some extent. Unfortunately, the process of fitting a large number of measurements with a function of a smaller number of parameters necessarily involves some loss of information. If we add *a priori* information to the problem by choosing a physically motivated fitting function, the information contained in the parameter error matrix may be entirely appropriate. To understand this problem better, we evaluate the uncertainty on $\Delta\alpha_{had}(M_Z^2)$ by two techniques. In the first, the parameter uncertainties are propagated to the calculated value of $\Delta\alpha_{had}(M_Z^2)$ using the following expression which is valid for any function of the parameters,

$$\delta^2(\Delta\alpha_{had})_{exp} = \sum_{k,l} \frac{\partial(\Delta\alpha_{had})}{\partial a_k} E_{kl} \frac{\partial(\Delta\alpha_{had})}{\partial a_l}, \qquad (14)$$

where the derivatives are calculated numerically and $E_{kl} = \langle \delta a_k \delta a_l \rangle$ is the parameter error matrix that is extracted from the fitting procedure. The second error estimate is performed by constructing a large ensemble of data sets by shifting the measured data points $R_{had}^i(\text{meas})$ as follows,

$$R_{had}^i(\text{set } j) = R_{had}^i(\text{meas}) + f_{ij}^{\text{ptp}}\sigma_i(\text{ptp}) + f_{ij}^{\text{norm}}\sigma_i(\text{norm}), \qquad (15)$$

where the factors $f_{ij}$ are Gaussian-distributed random numbers of unit variance. The entire fitting and integration procedure is then applied to each member of the ensemble. The uncertainty on $\Delta\alpha_{had}(M_Z^2)$ is determined from the central 68.3% of the ensemble distribution.

The use of a fitting function has the problem that one may introduce bias through the choice of parameterization. We attempt to evaluate this effect by varying the parameterizations as much as ingenuity and computer time allow. The quoted contributions to $\Delta\alpha_{had}(M_Z^2)$ are those corresponding to the best fits. Each contribution is assigned a parameterization uncertainty $\delta(\Delta\alpha_{had})_{param}$ based upon the spread of results corresponding to reasonable fits.

The first step in the evaluation of equation (5) for the hadronic continuum is to formulate a suitable (piecewise-continuous) parameterization $R_{fit}(s; a_k)$. We choose to use the perturbative QCD expression given in equation (9) in the region $W \geq 15$ GeV and an empirical parameterization in the region 1 GeV$\leq W <$ 15 GeV. In the high energy region, the only free parameter is $\alpha_s(M_Z^2)$ which is evolved to other scales using the prescription given by Marciano.[42]

In the portions of the low energy region that are measured well, polynomials are used to parameterize $R_{had}(W)$. To ensure that the function is continuous across several points $W_a$, the polynomials are constructed in $x_a \equiv W - W_a$ and the zeroth order terms are excluded,

$$P_n^a(x) \equiv \sum_{i=1}^{n} d_i^a x^i, \qquad (16)$$

where $a$ is a label to distinguish different regions. Separate polynomials are used to describe the following regions: 1 GeV$\leq W \leq 1.9$ GeV (labelled region s), 1.9 GeV$< W \leq 3.6$ GeV (labelled region c), and 5.0 GeV$< W \leq 10.4$ GeV (labelled region b). Although a single, large-order polynomial is adequate to describe the data between $W = 1$ GeV and charm threshold at 3.6 GeV, the data show a distinct shape change near $W = 1.9$ GeV (where the four-pion cross section is becoming small). It was possible to obtain better fits by introducing an additional polynomial to describe the region from 1 GeV to 1.9 GeV. A comparison of the two possible forms is used to assess the parameterization sensitivity of the final result.

Since there are no measurements of the continuum $R_{had}$ in the b-quark and c-quark threshold regions, it is necessary to extrapolate the form of $R_{had}$ from 3.6 GeV to 5.0 GeV and from 10.4 GeV to 15 GeV with functions that are physically motivated. In the case of the charm threshold region, the DASP Collaboration has published (in graphical form) the shape of the continuum that was preferred by their resonance fits. The function which characterizes the shape of the thresh-old, $f_{DASP}(W)$, does not increase as sharply as the free-quark threshold factor $\beta(3 - \beta^2)/2$ but increases more rapidly than the $\beta^3$ threshold factor for pointlike scalar particles. To construct the function $R_{fit}$, all three possibilities are used for the c-quark threshold and the two extreme possibilities are used for the b-quark threshold,

$$f_c(W) = \begin{cases} \beta(3 - \beta^2)/2 \\ f_{DASP}(W) \\ \beta^3 \end{cases} \qquad f_b(W) = \begin{cases} \beta(3 - \beta^2)/2 \\ \beta^3, \end{cases} \tag{17}$$

where the c- and b-quark masses are taken to be the $D$ and $B$ meson masses, respectively. The actual size of the charm-associated step in $R_{had}$, $\Delta R_c$ is left as a free parameter. The size of the bottom-associated step in $R_{had}$ is constrained to be the difference between the value of the fit function at $W = 10.4$ GeV and the value of the QCD portion at $W = 15$ GeV, $\Delta R_b \equiv R_{QCD}(15) - R_{fit}(10.4)$.

The actual form of the fitting function is given by the following expression,

$$R_{fit}(W) = \begin{cases} R_0 + P_{N_s}^s(W - 1.0), & 1 \leq W \leq 1.9 \\ R_{fit}(1.9) + P_{N_c}^c(W - 1.9), & 1.9 < W \leq 3.6 \\ R_{fit}(3.6) + \Delta R_c f_c(W), & 3.6 < W \leq 5.0 \\ R_{fit}(5.0) + P_{N_b}^b(W - 5.0), & 5.0 < W \leq 10.4 \\ R_{fit}(10.4) + \Delta R_b f_b(W), & 10.4 < W < 15.0 \\ R_{QCD}(W), & 15 \leq W \end{cases} \tag{18}$$

where $R_0$, the value of $R_{had}$ at $W = 1$ GeV, was a free parameter and the order of the polynomials varied from 1 to 7. The number of free parameters varied from 7 to 30. The fit quality did not improve substantially when the number of parameters exceeded 10. The weight matrix was constructed from equation (13) assuming that normalization uncertainties were completely correlated in four groups: the 20% uncertainties of the lowest energy measurements[21−22] (1.0 GeV< $W$ < 3.09 GeV), the 15-20% uncertainties of the MARK I, DASP, and PLUTO measurements[23−25] (2.6 GeV< $W$ < 4.9 GeV), the 5-10% uncertainties of the measurements[26−30] between charm and bottom thresholds, and the 1.7-7.0% uncertainties of the PEP and PETRA experiments[34-39] above bottom threshold. The data are corrected for electroweak interference before the fitting procedure is applied. The data and the result of a typical fit are shown in Fig. 2. The error bars include the point-to-point and the normalization uncertainties. The fit quality was excellent ($\chi^2/\text{dof} = 104.7/105$).

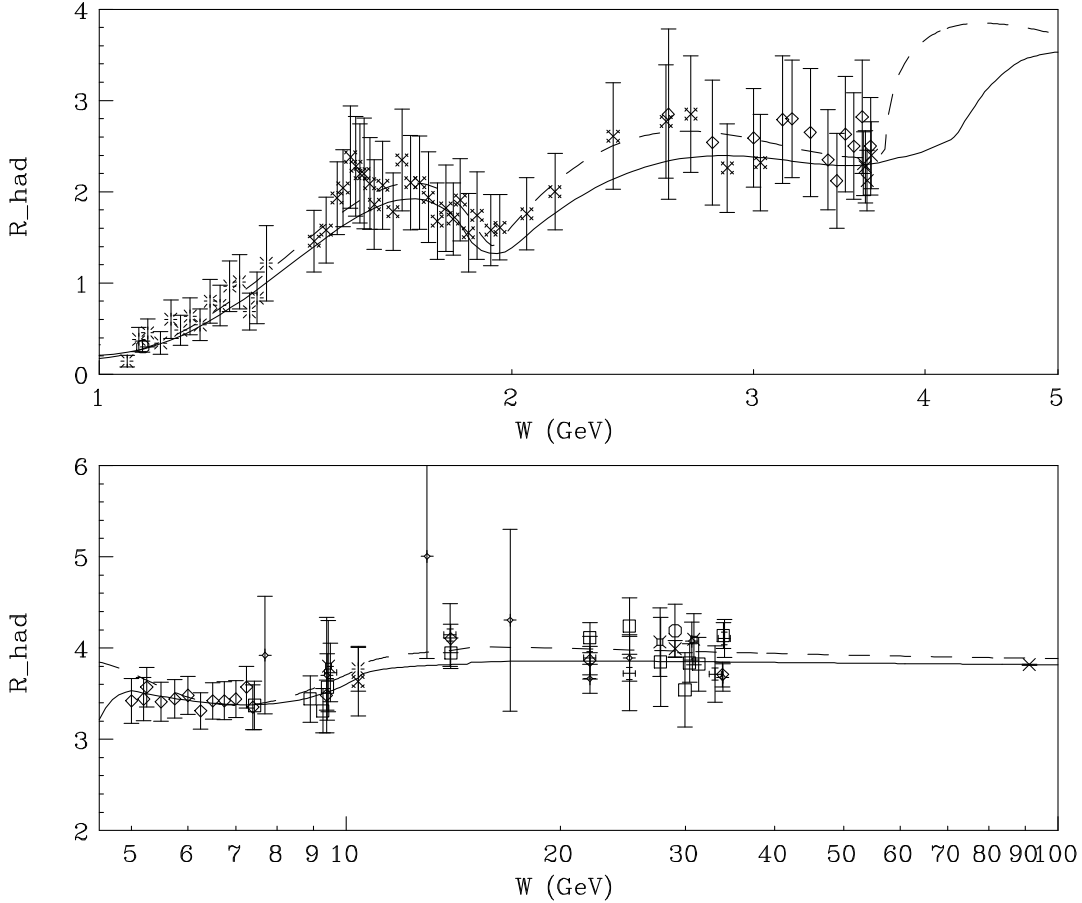

Figure 2. The continuum $R_{had}$ measurements including normalization uncertainties. A typical fit to equation (18) is shown as the solid curve. The dashed curve corresponds to the equal-weighting test described in the text.

The various hypotheses for $R_{fit}$ are used to evaluate the integral in equation (5) from $s_0 = 1$ GeV$^2$ to $\infty = 10^6$ GeV$^2$. Although the singularity in the integrand is formally well controlled, digital computers are famous for their inability to under-stand formalities. We have therefore recast equation (5) into a form which is more suitable for electronic evaluation,

$$\Delta\alpha_{had}(q^2) = \frac{\alpha_0 q^2}{3\pi}\left\{\frac{R_{fit}(q^2)}{q^2}\ln\left[\frac{q^2-s_0}{s_0}\right] - \int_{s_0}^{q^2-\Delta} ds\frac{R_{fit}(s)-R_{fit}(q^2)}{s(s-q^2)}\right.$$
$$\left.- \frac{\partial R_{fit}}{\partial s}\Big|_{q^2}\ln\left[\frac{q^2+\Delta}{q^2-\Delta}\right] - \int_{q^2+\Delta}^{\infty} ds\frac{R_{fit}(s)-R_{fit}(q^2)}{s(s-q^2)}\right\},$$

$$(19)$$

where we have assumed that $R_{fit}$ is well approximated by a linear expansion over the interval $q^2-\Delta < s < q^2+\Delta$ (in practice, we use $\Delta = 0.5$ GeV$^2$). The evaluation of equation (19) requires that $R_{QCD}(s)$ be extrapolated through t-quark threshold. For this purpose, the top quark mass is assumed to be 175 GeV.[43]

The contribution of the hadronic continuum to $\Delta\alpha_{had}(M_Z^2)$ is found to be fairly insensitive to the form of $R_{fit}$. The central value of $\Delta\alpha_{had}(M_Z^2)$ corresponds to the best estimate of the parameters of the function which uses: the DASP shape for the c-quark-threshold, the free-quark shape for the b-quark-threshold, and the values (2,3,3) for $(N_b, N_c, N_s)$. The maximum deviation from this value occurs when the $\beta^3$ function is used for the c-quark-threshold and a sixth-order polynomial is used to parameterize the entire region $W \leq 3.6$ GeV. The size of the maximum deviation is taken as an estimate of the parameterization uncertainty. The experimental uncertainty given by equation (14) is found to be a smaller by a factor of three than the estimate derived from the ensemble of fluctuated data sets. Since our fitting function has no physical motivation whatsoever in the low $W$ region, we choose the larger estimate as the more accurate. The resulting contribution to $\Delta\alpha_{had}(M_Z^2)$ is

$$\Delta\alpha_{had}^{\text{cont}}(M_Z^2) = 0.021428 \pm 0.000724(\text{exp}) \pm 0.000150(\text{param}). \qquad (20)$$

The experimental uncertainty given in equation (20) dominates the uncertainty on $\Delta\alpha_{had}(M_Z^2)$. We note that this uncertainty is itself dominated by the 0.000655 contribution of the normalization uncertainties of the $R_{had}$ measurements below charm threshold (particularly in the region 2.6-3.6 GeV). Any further improvement in the uncertainty on $\Delta\alpha_{had}(M_Z^2)$ requires that improved measurements be performed in this region.

The central value of this result is somewhat smaller than the one given in Ref. 4 for four reasons. The first is that we've defined the continuum contribution to exclude the charm-related enhancement near 4 GeV. The charm-threshold related enhancement is incorporated by the inclusion of the $\psi(4040)$, $\psi(4160)$, and $\psi(4415)$ resonances in the resonance contribution. The second difference is that our technique weights input data by their uncertainties and accounts for the large correlated uncertainties between the measurements within a measurement group. The third difference is that we have replaced the 44 MARK I measurements of the continuum between $W = 4.9$ GeV and $W = 7.6$ GeV with the more recent Crystal Ball data. The fourth difference is that we use the LEP measurement of $\alpha_s(M_Z^2)$ to constrain $R_{had}$ in the high energy region to a somewhat smaller value than the one preferred by the PEP/PETRA experiments alone. To verify that these differences are indeed source of the discrepancy, we have repeated the analysis with all points weighted equally (all points have constant fractional uncertainties and are assumed to be uncorrelated). To ensure that the sample is a good approximation of the one used in Ref. 4, all of the MARK I data including those measurements of the charm-threshold region are included and the Crystal Ball data and LEP measurement of $\alpha_s(M_Z^2)$ are excluded. The resulting value of $\Delta\alpha_{had}^{\rm cont}$, 0.0231, is close to the value given in Ref. 4 of 0.0233±0.0009.

The updated result given in Ref. 5 was derived by excluding the high energy MARK I data and including the Crystal Ball measurements. We simulate this result by substituting the Crystal Ball data for the 44 high energy MARK I measurements in the equally-weighted analysis (the charm threshold data are still included and the LEP measurement of $\alpha_s(M_Z^2)$ is still excluded). The resulting value of $\Delta\alpha_{had}^{\rm cont}$, 0.0226, is close to the value given in the second publication of Ref. 5 of 0.0228±0.0009. The best estimate of $R_{fit}$ resulting from the equally-weighted fit is shown as the dashed curve in Fig. 2. In addition to the difference in the charm threshold region, the equally-weighted fit prefers larger $R_{had}$ values in the region $W = 1$-3.6 GeV and in the region $W > 10$ GeV. In the $W = 2.6$-3.6 GeV region, the 15% normalization uncertainties of the DASP and PLUTO measurements pull the correctly-weighted fit to smaller $R_{had}$ values than those preferred by the 20% MARK I and $\gamma\gamma 2$ measurements. Since the $\gamma\gamma 2$ data extend to 1.42 GeV and are correlated by the large normalization uncertainty, the correctly-weighted fit is pulled to smaller $R_{had}$ values in the 1-2.6 GeV region. The use of the precise determination of $R_{had}$ at $W = M_Z$ is responsible for the $R_{fit}$ difference in the high energy region. The $R_{fit}$ differences in the $W$ regions 1-3.6 GeV, 3.6-6.0 GeV, and 8.0 GeV-$\infty$ lead to $\Delta\alpha_{had}$ differences of 0.0003, 0.0004, and 0.0005, respectively. The contribution of the three $\psi$ resonances in the charm threshold region is discussed in Section 2.4 and is found to be 0.0002, approximately one half of the discrepancy in the 3.6-6.0 GeV region. We conclude that we have identified the largest part of the discrepancy and have validated our technique.

The experimental uncertainty given in equation (20) is also smaller than the corresponding one given in Ref. 4. The authors of Ref. 4 evaluated their uncertainty by assigning normalization uncertainties in several $W$ regions and adding them in quadrature. They assessed the following uncertainties: 20% in the region 1 GeV$<$ $W < 2.3$ GeV, 10% in the region 2.3 GeV$< W < 12$ GeV, and 3% in the region $W > 12$ GeV. Note that the normalization uncertainty changes discontinuously at the boundaries of each region. The large uncertainties at small $W$ are moderated by the fact that most of the integral accrues in the high energy region. Our use of $R_{had}(M_Z)$ essentially eliminates the experimental uncertainty in the region $W > 15$ GeV. The Crystal Ball data constrain the uncertainty in the region between c-quark threshold and b-quark threshold to be approximately one half of that assumed in Ref. 4. Below charm threshold, the normalization uncertainty increases from 15% near 3 GeV to 20% below 2 GeV. Since we require that $R_{fit}$ be a continuous (and reasonably smooth) function, the normalization uncertainty is required to vary smoothly as $W$ decreases. The better measured larger $W$ regions constrain the uncertainty at the smaller $W$ regions.

## 2.3 THE $\pi^+\pi^-$ AND $K^+K^-$ FINAL STATES

The processes $e^+e^- \to \pi^+\pi^-$ and $e^+e^- \to K^+K^-$ are described by the electromagnetic form factors, $F_\pi(s)$ and $F_K(s)$, which are related to the hadronic cross section ratio $R_{had}$ for each process as follows,

$$R_{had}^{\pi^+\pi^-}(s) = \frac{1}{4}|F_\pi(s)|^2\beta_\pi^3, \qquad R_{had}^{K^+K^-}(s) = \frac{1}{4}|F_K(s)|^2\beta_K^3, \qquad (21)$$

where $\beta_\pi$ and $\beta_K$ are the velocities of the final state particles in the $e^+e^-$ center-of-mass frame. It is clear that measurements of the form factors are equivalent to measurements of $R_{had}$.

Measurements of the square of the pion form factor $|F_\pi|^2$ have been performed by the OLYA,[7] CMD,[7] TOF,[9] NA7,[8] $\mu\pi$,[12] MEA,[14] M2N,[10] DM1,[11] and DM2[13] Collaborations and are shown in Fig. 3. The error bars include the normalization uncertainties which range from about 2% in the region around the (dominant) $\rho$ resonance to about 12% at $W \simeq 2$ GeV.

The data are fit to a function which is a sum of the Gounaris-Sakurai form[44] used by Kinoshita, Nizic, and Okamoto[45] and three resonances,

$$F_\pi(s) = \frac{A_1 - m_\pi^2 A_2}{A_1 + A_2 q^2 + f(s)} + \sum_{n=1}^{3} \frac{B_n e^{iC_n} m_n^2}{s - m_n^2 + im_n\Gamma_n}, \qquad (22)$$

where: $A_1$ and $A_2$ are free parameters; $m_\pi$ is the pion mass; $q$ and $f(s)$ are defined as follows,

$$\begin{aligned} q &\equiv \sqrt{s/4 - m_\pi^2} \\ f(s) &\equiv \frac{1}{\pi}\left[m_\pi^2 - \frac{s}{3}\right] + \frac{2q^3}{\pi\sqrt{s}}\ln\left[\frac{\sqrt{s} + 2q}{2m_\pi}\right] - i\frac{q^3}{\sqrt{s}}; \end{aligned} \qquad (23)$$

and where $m_n$, $\Gamma_n$, $B_n$, and $C_n$ are the mass, width, amplitude, and phase of each resonance. The mass and width of the first resonance were set to those of the $\omega(782)$. All other parameters (12 in total) were allowed to vary. The weight
matrix of the fit was constructed by assuming that all normalization uncertainties
are 100% correlated (see equation (13)). The result of the fit is shown as a solid
line in Fig. 3. The fit preferred a resonance of width 0.36 GeV at mass 1.2 GeV
and a second resonance of width 0.16 GeV at mass 1.7 GeV. The fit quality is
found to be excellent ($\chi^2/\text{dof} = 116.7/128$).

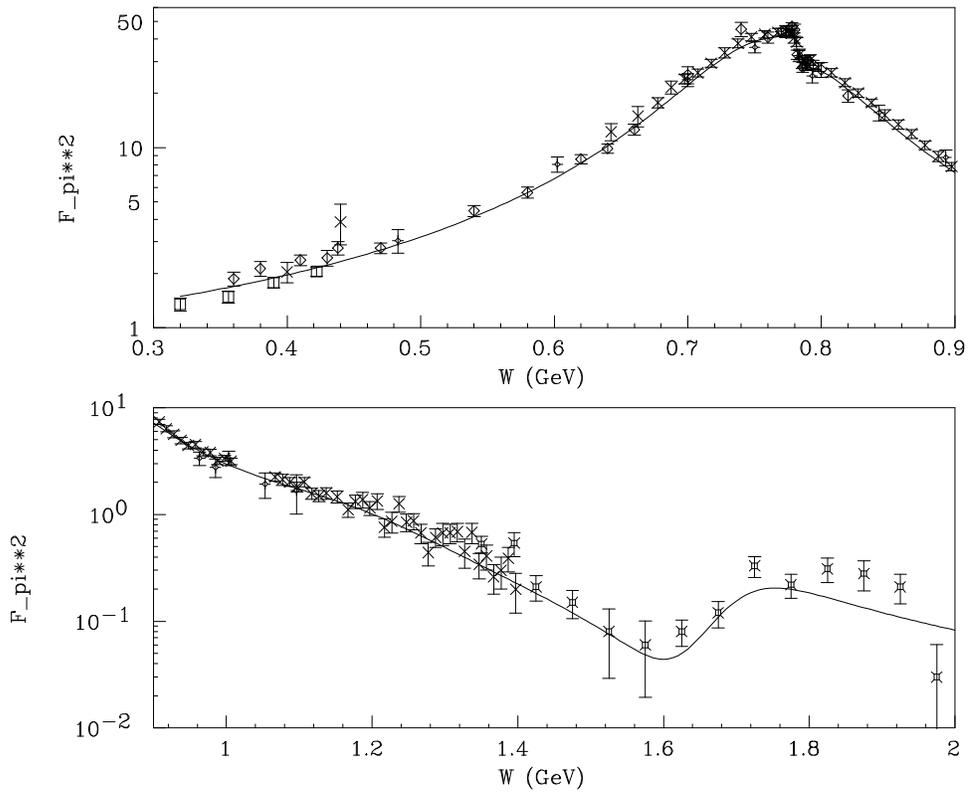

Figure 3. Measurements of $|F_\pi(W)|^2$ are compared with the best fit which is shown
as a solid line. The error bars include normalization uncertainties.

To evaluate the sensitivity of the result to the parameterization, the complete
function used by the authors of Ref. 45 was also fit to the data. This function did
not fit the newest data from DM2 (at large $W$) as well as our chosen form. Both
functions were used to evaluate equation (5) from $s = 4m_\pi^2$ to $s = 4$ GeV$^2$ (where
$|F_\pi|^2$ is measured to be very small). We find the $\pi^+\pi^-$ contribution to $\Delta\alpha_{had}(M_Z^2)$

to be

$$\Delta\alpha_{had}^{\pi^+\pi^-}(M_Z^2) = 0.003087 \pm 0.000051(\text{exp}) \pm 0.000121(\text{param}). \qquad (24)$$

The two techniques for the estimation of the experimental uncertainty (discussed in Section 2.2) yielded consistent results in this case.

The result given in equation (24) differs from the corresponding result given in Ref. 4 mostly because of the inclusion of the large-$W$ DM2 data which decrease more sharply with energy than the extrapolated tail of the function that was fit to the lower-energy data. The uncertainty quoted by the authors of Ref. 4 corresponds to the fractional uncertainty on the leptonic width of the $\rho$.

Measurements of the square of the kaon form factor $|F_K|^2$ have been performed by the OLYA,[15] CMD,[16] MEA,[14] DM1,[17] and DM2[18] Collaborations and are shown in Fig. 4. The data span the $\phi(1020)$ resonance and continue to $W = 1.8$ GeV where $R_{had}^{K^+K^-}$ is less than 0.01. The normalization uncertainty on the CMD measurements is 6%. The other groups do not report normalization uncertainties. Early $|F_\pi|^2$ measurements suffered from the same problem of unreported normalization uncertainties. A bit of historical research shows that the normalization uncertainties were usually not included in the point-to-point errors. We therefore arbitrarily assign a 20% systematic normalization uncertainty to all unreported cases. The data and total uncertainties are shown in Fig. 4.

The data are fit to a function which is a sum of a Breit-Wigner resonance with an energy-dependent width for the $\phi$ and four resonances,

$$F_K(s) = \frac{A_1}{s - m_\phi + im_\phi\Gamma_\phi(s)} + \sum_{n=1}^{4} \frac{B_n e^{iC_n}}{s - m_n^2 + im_n\Gamma_n}, \qquad (25)$$

where: $A_1$ is the amplitude of the $\phi$; $m_\phi$ is the mass of the $\phi(1020)$; $m_n$, $\Gamma_n$, $B_n$, and $C_n$ are the mass, width, amplitude, and phase of the resonances. The energy-dependent width $\Gamma_\phi(s)$ is assumed to consist of contributions from the $K^+K^-$,

$K_L K_S$, and $3\pi$ final states,

$$\Gamma_\phi(s) = \Gamma_\phi^0 \left\{ \frac{\sqrt{s}}{m_\phi} \left[ 0.497 \frac{\beta_+^3(s)}{\beta_+^3(m_\phi^2)} + 0.347 \frac{\beta_0^3(s)}{\beta_0^3(m_\phi^2)} \right] + 0.156 G_{3\pi}^\phi(s) \right\}, \qquad (26)$$

where: $\Gamma_\phi^0$ is the nominal value[6] of the $\phi$ width, $\beta_+(s) = \sqrt{1 - 4m_{K^+}^2/s}$ is the velocity of the charged kaon, $\beta_0(s) = \sqrt{1 - 4m_{K^0}^2/s}$ is the velocity of the neutral kaon, and $G_{3\pi}^\phi(s)$ is a function which is normalized to unity at $s = m_\phi^2$ and is proportional to the decay rate for $\phi \to 3\pi$ assuming $\rho\pi$ dominance.[46]

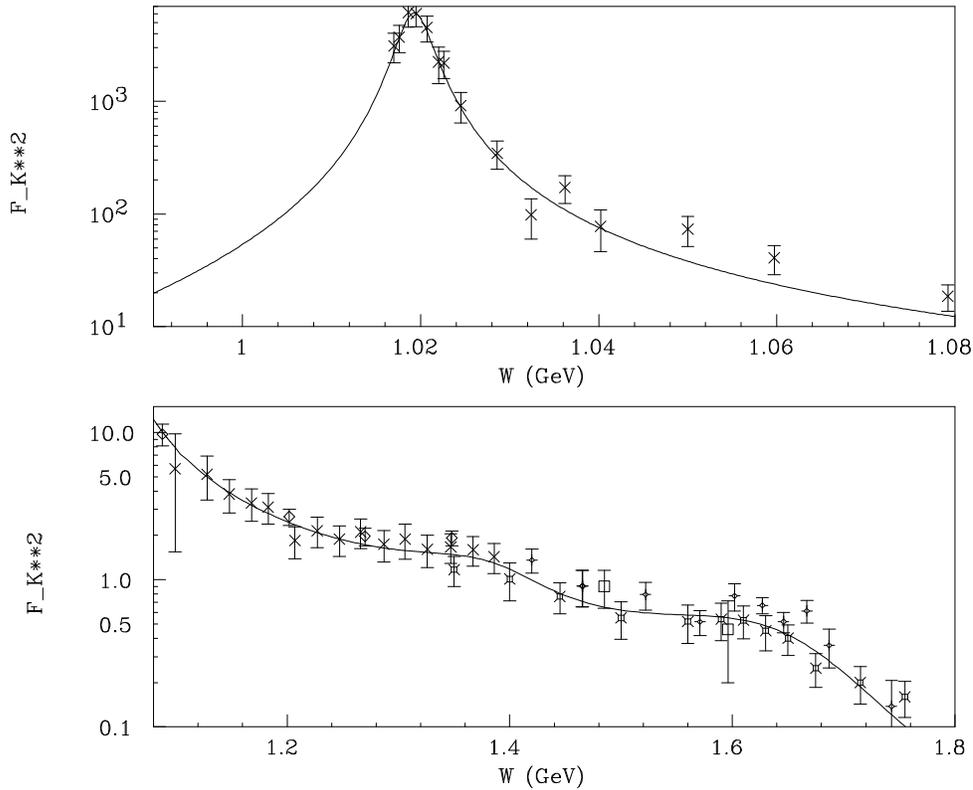

Figure 4. Measurements of $|F_K(W)|^2$ are compared with the best fit which is shown as a solid line. The error bars include normalization uncertainties.

The masses and widths of the first two resonances were set to those of the $\rho(770)$ and $\omega(782)$. Following the procedure of Ref. 18, the amplitude ratios $B_1/A_1$ and $B_2/A_1$ were constrained to the measured values and the phases were set to zero. The mass, width, and amplitude of the $\phi$ were allowed to vary. The masses, widths, amplitudes, and phases of two larger mass resonances were free parameters. The weight matrix of the fit was constructed by assuming that all normalization uncertainties are 100% correlated (see equation (13)). The result of the fit is shown as a solid line in Fig. 4. The fit preferred a resonance of width 0.15 GeV at mass 1.39 GeV and a second resonance of width 0.22 GeV at mass 1.65 GeV. The fit quality is found to be adequate ($\chi^2/\mathrm{dof} = 74.1/49$).

To evaluate the sensitivity of the result to the parameterization, a second fit was performed with the amplitudes and phases of the $\rho$ and $\omega$ allowed to vary as free parameters. The fit quality improved marginally ($\chi^2/\mathrm{dof} = 69.6/45$). Both functions were used to evaluate equation (5) from $s = 4m_{K^+}^2$ to $s = 3.24 \text{ GeV}^2$. We find the $K^+K^-$ contribution to $\Delta\alpha_{had}(M_Z^2)$ to be

$$\Delta\alpha_{had}^{K^+K^-}(M_Z^2) = 0.000311 \pm 0.000030(\mathrm{exp}) \pm 0.000009(\mathrm{param}). \qquad (27)$$

The two techniques for the estimation of the experimental uncertainty (discussed in Section 2.2) also yielded consistent results in this case. To account for the poor fit quality, the experimental uncertainty has been scaled by the factor 1.23.

## 2.4 THE RESONANCES

The resonances comprise the remaining portion of the total $e^+e^-$ cross section. The total cross section for each resonance can be represented by a relativistic Breit-Wigner form with energy-dependent widths,

$$\sigma_{res}(s) = \frac{12\pi}{m^2} \frac{s\Gamma_{ee}(s)\Gamma_{tot}(s)}{(s-m^2)^2 + s\Gamma_{tot}^2(s)}, \qquad (28)$$

where: $m$, $\Gamma_{ee}$, and $\Gamma_{tot}$ are the mass, electronic width, and total width of the resonance. In order to incorporate the Breit-Wigner cross sections described by equation (28) into equation (5), it must be scaled to the electromagnetic point cross section, $\sigma_{\mu\mu}(s) = 4\pi\alpha^2(s)/3s$, yielding the following expression,

$$\Delta\alpha_{had}^{res}(q^2) = \frac{\alpha_0 q^2}{4\pi^2} \mathrm{P} \int\limits_{4m_\pi^2}^{\infty} ds \frac{\sigma_{res}(s)}{\alpha^2(s)[q^2 - s]}, \tag{29}$$

which has the slightly unpleasant feature that it incorporates $\alpha(s)$, the quantity that we are attempting to evaluate, into the integrand. Equation (29) is often written in the approximation, $\alpha(s) \simeq \alpha_0$. Unfortunately, this overestimates $\Delta\alpha_{had}(M_Z^2)$ by: 3% at the $\omega(783)$, 5% at the $J/\psi(1S)$, and 7% at the $\Upsilon(1S)$. To avoid this problem, we use the $\Delta\alpha_{had}(s)$ parameterization given in Ref. 4 to generate a first-order estimate of $\alpha(s)$ for use in equation (29).

Equation (29) is evaluated for the $\omega(782)$, $\phi(1020)$, $\psi$-family, and $\Upsilon$-family resonances by performing a Simpson's rule integration over the interval $m - 60\Gamma_{tot}$ to $m + 60\Gamma tot$ (the lower limit of the $\omega$ integration is the threshold for $3\pi$ decay). The correction for electromagnetic decays is performed using the best estimate of the function $R_{fit}(s)$ determined in Section 2.3. The energy-dependent electronic widths and the hadronic widths of the $\psi$ and $\Upsilon$ resonances are assumed to scale as $\sqrt{s}$,

$$\Gamma(s) = \frac{\sqrt{s}}{m}\Gamma_0, \tag{30}$$

where $m$ is the mass of the resonance and $\Gamma_0$ is the nominal value of the width. The energy-dependent total hadronic width of the $\phi(1020)$ is given by equation (26). The width $\Gamma'_{had}$ for the $\phi$ is adjusted to exclude the $K^+K^-$ final state (discussed in Section 2.4). The energy-dependent total hadronic width of the $\omega(782)$ is given by the following expression which assumes that all final states are $\pi^+\pi^-$, $\pi^0\gamma$, or $\pi^+\pi^-\pi^0$,

$$\Gamma_\omega(s) = \Gamma_\omega^0 \left\{ \frac{\sqrt{s}}{m_\omega} \left[ 0.022\frac{\beta_\pi^3(s)}{\beta_\pi^3(m_\omega^2)} + 0.085\frac{(1 - m_\pi^2/s)^3}{(1 - m_\pi^2/m_\omega^2)^3} \right] + 0.893 G_{3\pi}^\omega(s) \right\}, \tag{31}$$

where: $m_\omega$ is the mass of the $\omega$, $\Gamma_\omega^0$ is the nominal value[6] of the $\omega$ width, $\beta_\pi(s) = \sqrt{1 - 4m_\pi^2/s}$ is the velocity of the charged pion, and $G_{3\pi}^\omega(s)$ is a function which is normalized to unity at $s = m_\omega^2$ and is proportional to the decay rate for $\omega \to 3\pi$ assuming a constant matrix element (phase space weighting).

The masses and widths used to evaluate equation (29) are taken from the 1994 Review of Particle Properties[6]. The results are listed in Table 1 along with those derived in Sections 2.3 and 2.4. The experimental uncertainties are evaluated by assuming that the uncertainties on the masses, total widths, electronic widths, and relevant branching ratios are uncorrelated. The parameterization uncertainties are evaluated by repeating the calculation with a constant-width, constant-mass Breit-Wigner cross section.

**Table 1**: Summary of the various contributions to $\Delta\alpha_{had}$.

| Contribution | W Region (GeV) | $\Delta\alpha_{had}(M_Z^2)$ | $\delta(\Delta\alpha_{had})_{exp}$ | $\delta(\Delta\alpha_{had})_{param}$ |
|---|---|---|---|---|
| Continuum | 1.0-$\infty$ | 0.021428 | 0.000724 | 0.000150 |
| $\pi^+\pi^-$ | 0.280-2.0 | 0.003087 | 0.000051 | 0.000121 |
| $K^+K^-$ | 0.987-1.8 | 0.000307 | 0.000025 | 0.000009 |
| Resonances | $\omega^{(a)}$ | 0.000305 | 0.000010 | 0.000003 |
| " | $\phi^{(b)}$ | 0.000304 | 0.000011 | 0.000004 |
| " | $\psi$ (6 states) | 0.001106 | 0.000059 | 0.000023 |
| " | $\Upsilon$ (6 states) | 0.000118 | 0.000005 | 0.000003 |
| Total | | 0.02666 | 0.00072 | 0.00019 |

[a]Doesn't include $\pi^+\pi^-$ final states.

[b]Doesn't include $K^+K^-$ final states.

The only resonance entry in Table 1 that is directly comparable to a corresponding result in Ref. 4 is $\Upsilon$-family result which agrees well despite the use of $\alpha^2(s)$ in equation (29). The sum of our $K^+K^-$ and $\phi$ entries is larger than the corresponding $\phi$ result of Ref. 4 by 16%. Taking the correction for $\alpha(s)$ into account, this implies that continuum $K^+K^-$ final states contribute about 20% of the $\phi$ contribution to $\Delta\alpha_{had}(M_Z^2)$. Our $\psi$-family result is larger than the result given in Ref. 4 by 2% even though it includes three additional states. As a cross check, a repetition of the calculation with $\Gamma'_{had} = \Gamma_{tot}$, $\alpha(s) = \alpha_0$, the constant-width and mass Breit-Wigner function, and 1988 values of the resonance parameters did agree very well with the numbers given in Ref. 4.

## 2.5 Final Result

The various contributions to $\Delta\alpha_{had}(M_Z^2)$ are summarized and summed in Table 1. The resulting value,

$$\Delta\alpha_{had}(M_Z^2) = 0.02666 \pm 0.00075, \tag{32}$$

differs by 2.6 (Ref. 4) standard deviations from the result given in Ref. 4 and by 1.9 standard deviations from the updated result given in Ref. 5. Including the leptonic contribution, we find $\alpha^{-1}(M_Z^2)$ to be,

$$\alpha^{-1}(M_Z^2) = 129.08 \pm 0.10, \tag{33}$$

where the uncertainties on the lepton masses contribute negligibly to the total uncertainty.

# 3. Interpretation

Since most electroweak calculations are based upon input parameters ($\alpha_0$, $M_Z$, and the Fermi coupling constant, $G_F$) that are measured at very different scales, the quantity $\alpha(M_Z^2)$ enters into the calculation of most electroweak observables. To understand the effect of our result upon the Standard Model predictions for various electroweak observables, we have used the value of $\alpha(M_Z^2)$ given in equation (33) with the ZFITTER 4.8 program of Bardin, *et al.*[47] to calculate: the mass of the $W$ boson ($M_W$), the width of the $Z$ boson ($\Gamma_Z$), the ratio of the hadronic decay width of the $Z$ to the (single species) leptonic decay width ($R_\ell$), the tree-level total hadronic cross section at the $Z$ pole ($\sigma_{had}^0$), the effective weak mixing angle at the $Z$ pole ($\sin^2 \theta_W^{\mathrm{eff}}$), the ratio of the $b\bar{b}$ decay width of the $Z$ to the hadronic width ($R_b$), the ratio of the neutral and charged current cross sections in neutrino nucleon scattering ($R_\nu$), and the weak charge of the Cesium nucleus as measured in atomic parity violation experiments ($Q_W(\mathrm{Cs})$). The default value for $\Delta\alpha_{had}(M_Z^2)$ in ZFITTER is the result of Ref. 4 with the result of Ref. 5 available as an option. The shifts in the observables at a top quark mass ($m_t$) of 175 GeV and a Higgs boson mass ($m_H$) of 300 GeV from those calculated with the default value of $\Delta\alpha_{had}(M_Z^2)$ are listed in Table 2. The shifts are also normalized to the current world-average experimental uncertainties on each quantity. It is clear that the interpretation of the current measurements of $\sin^2 \theta_W^{\mathrm{eff}}$ is most affected and that the interpretation of the $\Gamma_Z$ and $R_\ell$ measurements is moderately affected.

**Table 2**: Shifts in the predicted values of various electroweak observables using the value of $\Delta\alpha_{had}(M_Z^2)$ given in equation (33) with the ZFITTER 4.8 program of Bardin *et al.*[47] The shifts are calculated at $m_t = 175$ GeV and $m_H = 300$ GeV.

| Observable | Shift wrt Ref. 4 [Ref. 5] | Shift/Exptl Error |
|:---:|:---:|:---:|
| $M_W$ | $+40$ [$+30$] MeV | $+0.22$ [$+0.17$] |
| $\Gamma_Z$ | $+1.9$ [$+1.4$] MeV | $+0.50$ [$+0.37$] |
| $R_\ell$ | $+0.013$ [$+0.010$] | $+0.33$ [$+0.25$] |
| $\sigma_{had}^0(Z)$ | $-0.004$ [$-0.003$] nb | $-0.04$ [$-0.03$] |
| $\sin^2 \theta_W^{\mathrm{eff}}$ | $-0.00074$ [$-0.00055$] | $-1.8$ [$-1.4$] |
| $R_b$ | $-0.00003$ [$-0.00002$] | $-0.015$ [$-0.010$] |
| $R_\nu(\mathrm{nc/cc})$ | $+0.0005$ [$+0.0004$] | $+0.18$ |
| $Q_W(\mathrm{Cs})$ | $+0.163$ [$+0.122$] | $+0.09$ [$+0.07$] |

The effect on the interpretation of $\sin^2 \theta_W^{\mathrm{eff}}$ can be made more clear by calculating the allowed range for the currently favored value[43] of the top quark mass, $m_t = 174$ GeV, as the Higgs boson mass varies from 60 GeV to 1 TeV,

$$\sin^2 \theta_W^{\mathrm{eff}} = \begin{cases} 0.2306, & m_H = 60 \text{ GeV} \\ 0.2315, & m_H = 300 \text{ GeV} \\ 0.2322, & m_H = 1 \text{ TeV}. \end{cases} \quad (34)$$

The current determination of $\sin^2 \theta_W^{\text{eff}}$ by the LEP Collaborations,[40] 0.2321±0.0004, is consistent with the CDF top mass value and a heavy Higgs boson. The value of $\sin^2 \theta_W^{\text{eff}}$ extracted from the left-right cross section asymmetry in $Z$ production by the SLD Collaboration,[48] 0.2294±0.0010, is smaller than the light Higgs value by 1.2 standard deviations.

**Table 3**: The results of global MSM and *S-T-U* fits to the electroweak observables listed in Table 2.

| Parameter | $\alpha^{-1}(M_Z^2)$ | | |
|---|---|---|---|
| | $128.80 \pm 0.12$ | $128.87 \pm 0.012$ | $129.08 \pm 0.10$ |
| Standard Model Fit | | | |
| $m_t$ (GeV) | $172.6^{+10.2+16.3}_{-10.8-18.1}$ | $169.7^{+10.3+16.4}_{-10.9-18.3}$ | $161.4^{+10.4+17.0}_{-11.0-19.4}$ |
| $\alpha_s$ | 0.125±0.005 | 0.125±0.005 | 0.125±0.005 |
| $\chi^2$/dof | 22.3/12 | 21.6/12 | 20.0/12 |
| *S-T-U* Fit | | | |
| $S$ | $-0.37\pm0.23$ | $-0.31\pm0.23$ | $-0.16\pm0.22$ |
| $T$ | $+0.27\pm0.23$ | $+0.27\pm0.23$ | $+0.28\pm0.23$ |
| $U$ | $-0.20\pm0.56$ | $-0.18\pm0.56$ | $-0.14\pm0.56$ |
| $\alpha_s$ | 0.124±0.005 | 0.124±0.005 | 0.123±0.005 |
| $\chi^2$/dof | 11.4/9 | 11.5/9 | 11.6/9 |

We have performed a fit of the ZFITTER model to the current best measurements[40] of all of the quantities listed in Table 2 (the seven different observables that determine $\sin^2 \theta_W^{\text{eff}}$ are entered separately). The fit was performed with the parameter $\alpha^{-1}(M_Z^2)$ constrained to 128.80±0.12 (corresponds to the value of $\Delta\alpha_{had}(M_Z^2)$ given in Ref. 4), 128.87±0.12 (corresponds to the value of $\Delta\alpha_{had}(M_Z^2)$ given in Ref. 5), and 129.08±0.10. The full correlation matrix for the LEP measurements is included in the fit. The top quark mass and the strong coupling constant were allowed to vary as free parameters in a series of three fits with the Higgs boson mass set to: 60 GeV, 300 GeV, and 1 TeV. The results are listed in Table 3 for the $m_H = 300$ GeV case. The extracted value of $m_t$ is sensitive to the choice of Higgs mass. The range of sensitivity is indicated by the second set of errors as $m_H$ is varied from 60 GeV (lower value) to 1 TeV (upper value). Note that the use of our new value of $\alpha(M_Z^2)$ decreases the extracted value of $m_t$ by 11 GeV [8 GeV] as compared with the result of Ref. 4 [Ref. 5] and slightly improves the quality of the fit.

We have also performed a fit of the Peskin-Takeuchi $S$-$T$-$U$ parameterization[49] to the best current measurements[40] of the quantities listed in Table 2. The parameters $S$, $T$, $U$, and $\alpha_s(M_Z^2)$ were allowed to vary as free parameters with $\alpha^{-1}(M_Z^2)$ constrained to 128.80±0.12, 128.87±0.12, and 129.08±0.10. The results are listed in Table 3 for the reference values of $m_t$ and $m_H$ taken to be 150 GeV and 1 TeV, respectively. The use of the new value of $\alpha(M_Z^2)$ shifts the extracted value of $S$ by +0.21 [+0.16] as compared with the result of Ref. 4 [Ref. 5].

## 4. Conclusions

We have reevaluated the hadronic part of the electromagnetic vacuum expectation value using the standard dispersion integral approach that utilizes the hadronic cross section measured in $e^+e^-$ experiments as input. Previous analyses are based upon point-by-point trapezoidal integration which has the effect of weighting all inputs equally. We use a technique that weights the experimental inputs by their stated uncertainties, includes correlations, and incorporates some refinements. We find the hadronic contribution to $\Delta\alpha(M_Z^2)$ to be,

$$\Delta\alpha_{had}(M_Z^2) = 0.02666 \pm 0.00075,$$

which leads to a value of the electromagnetic coupling constant at $s = M_Z^2$,

$$\alpha^{-1}(M_Z^2) = 129.08 \pm 0.10.$$

Our value of $\alpha(M_Z^2)$ shifts the predicted values of a number of electroweak observables. The most affected is the effective weak mixing angle at the $Z$ pole which shifts by $-0.0007$ [$-0.0006$] from that predicted using the $\Delta\alpha_{had}(M_Z^2)$ value of Ref. 4 [Ref. 5]. The best estimate of the top quark mass extracted from a global fit shifts by $-11$ GeV [$-8$ GeV] and the best estimate of the Peskin-Takeuchi $S$ parameter shifts by $+0.21$ [$+0.16$].

We note that the current generation of $\sin^2\theta_W^{\text{eff}}$ measurements are likely to saturate the $\pm 0.00026$ uncertainty due to the $\pm 0.10$ uncertainty on $\alpha^{-1}(M_Z^2)$. The best hope for improvement is for the BES Collaboration to make improved measurements of $R_{had}$ in the region $W = 2 - 3.6$ GeV. A modest set of measurements in this region with a normalization uncertainty $\delta R_{had}/R_{had} \lesssim 5\%$ would reduce the current uncertainty by a factor of two and would eliminate it as a limitation.

Acknowledgements:

This work was originally motivated by the Long Term Planning Study organized by the Division of Particles and Fields of the American Physical Society. As part of his attempt to penetrate the popular myths pertaining to the uncertainty on $\alpha(M_Z^2)$, the author approached Michael Peskin who correctly pointed out that the only way to really understand the issues was to repeat the analysis. The author would like to thank Michael for his initial suggestion, his useful discussions and technical advice, and his comments on this manuscript. This work was supported by Department of Energy Contract No. DE-AC03-76SF00515.